\documentclass[a4paper,11pt]{article}
\pdfoutput=1 

\usepackage{jcappub} 

\usepackage[T1]{fontenc} 

\title{\boldmath Searching for primordial non-Gaussianity in Planck CMB maps 
        using a combined estimator}

\author[a]{C. P. Novaes,}
\author[b]{A. Bernui,}
\author[c]{I. S. Ferreira}
\author[a]{and C. A. Wuensche}

\affiliation[a]{Divis\~ao de Astrof\'isica, Instituto Nacional de Pesquisas Espaciais,\\Av. dos Astronautas 1758, S\~ao Jos\'e dos Campos 12227-010, SP, Brazil}
\affiliation[b]{Observat\'orio Nacional,\\Rua General Jos\'e Cristino 77, 
          S\~ao Crist\'ov\~ao, 20921-400, Rio de Janeiro, RJ, Brazil}
\affiliation[c]{Instituto de F\'{\i}sica, Universidade de Bras\'{\i}lia, 
Campus Universit\'ario Darcy Ribeiro,\\Asa Norte, 70919-970, Bras\'{\i}lia, DF, Brazil}

\emailAdd{camilapnovaes@gmail.com}
\emailAdd{bernui@on.br}
\emailAdd{ivan@fis.unb.br}
\emailAdd{ca.wuensche@inpe.br}

\abstract{The extensive search for deviations from Gaussianity in cosmic microwave background radiation (CMB) data is very important due to the information about the very early moments of the universe encoded there. Recent analyses from {\it Planck} CMB data do not exclude the presence of non-Gaussianity of small amplitude, although they are consistent with the Gaussian hypothesis. The use of dif\/ferent techniques is essential to provide information about types and amplitudes of non-Gaussianities in the CMB data. In particular, we find interesting to construct an estimator based upon the combination of two powerful statistical tools that appears to be sensitive enough to detect tiny deviations from Gaussianity in CMB maps. This estimator combines the Minkowski functionals with a Neural Network, maximizing a tool widely used to study non-Gaussian signals with a reinforcement of another tool designed to identify patterns in a data set. We test our estimator by analyzing simulated CMB maps contaminated with dif\/ferent amounts of {\em local} primordial non-Gaussianity quantified by the dimensionless parameter $f_{\rm \,NL}$. We apply it to these sets of CMB maps and find $\gtrsim$ 98\% of chance of positive detection, even for small intensity local non-Gaussianity like $f_{\rm \,NL} = 38 \pm 18$, the current limit from {\it Planck} data for large angular scales. Additionally, we test the suitability to distinguish between primary and secondary non-Gaussianities: first we train the Neural Network with two sets, one of nearly Gaussian CMB maps ($|f_{\rm \,NL}| \le 10$) but contaminated with realistic inhomogeneous {\it Planck} noise (i.e., secondary non-Gaussianity) and the other of non-Gaussian CMB maps, that is, maps endowed with weak primordial non-Gaussianity ($28 \le f_{\rm \,NL} \le 48$); after that we test an ensemble composed of CMB maps either with one of these non-Gaussian contaminations, and find out that our method successfully classifies $\sim 95 \%$ of the tested maps as being CMB maps containing primordial or secondary non-Gaussianity. Furthermore, we analyze the foreground-cleaned {\it Planck} maps obtaining constraints for non-Gaussianity at large-angles that are in good agreement with recent constraints. Finally, we also test the robustness of our estimator including cut-sky masks and realistic noise maps measured by {\it Planck}, obtaining successful results as well.
}

\begin{document}
\maketitle
\flushbottom

\section{Introduction}
\label{Introduction}

The spectacular advance in modern observational cosmology is mainly due, on the one side, 
to the highly sensitive measurements of the cosmic microwave background radiation (CMB) 
performed by the Wilkinson Microwave Anisotropy Probe 
(WMAP)~\citep{WMAP1a,WMAP3a,WMAP5a,WMAP7a,WMAP9a} 
and the {\it Planck} satellite~\citep{PLA-I,PLA-XV}, and, on the other side, to the several large and 
deep galaxy surveys, like the 2dFGRS~\citep{2dFGRS}, 6dFGRS~\citep{6dFGRS}, 
2MASS~\citep{2MASS} and SDSS~\citep{SDSS} projects, which mapped the luminous large-scale 
matter distribution. 
However, while large scale massive structures give information regarding more recent times 
of the universe evolution, at redshifts up to $z \sim 2$, CMB is the oldest cosmological observable, 
at $z \sim 1100$, with present-day technologies. 
The statistical properties of the CMB temperature f\/luctuations provides unique 
cosmological information from physical processes occurred in the early universe, such as 
those that may have produced the primordial density perturbations that evolved 
gravitationally towards the presently observed large scale structures (see, 
e.g.,~\citep{Bartolo10,Bartolo10/2,Komatsu1}). 

In the standard model of cosmology, inf\/lation is considered the dominant paradigm for 
the generation of such primordial density perturbations regarded as the seeds for structure 
formation.
The analysis of Gaussian deviations in the CMB temperature f\/luctuations is considered a powerful 
probe to investigate the nature of these perturbations
~\citep{Andrade,Abramo09,Golovnev,Kawasaki2,Kawasaki3,%
Komatsu1,Komatsu2,Abramo10,Chen,Sasaki,Valenzuela-Toledo,Pogosyan}. 
Present and forthcoming CMB data seems to be sensitive enough to detect tiny Gaussian 
deviations of the type and amplitude expected to be generated by small non-linear 
density perturbations~\citep{Komatsu02,Komatsu2}. 
In particular,  the recently released {\it Planck} data~\citep{PLA-I} 
are considered one the most precise data sets to investigate the physics of the very early 
universe.

The detection of primordial non-Gaussianity (NG) in the CMB data is crucial to 
discriminate among inf\/lationary models and also to test alternative 
scenarios (see, e.g.,~\citep{Bartolo04,Komatsu1,Chen,Bartolo10,Komatsu2}). 
Since non-cosmological ef\/fects can introduce non-Gaussian contaminations in CMB 
maps, the discrimination of a \emph{primordial} non-Gaussian signal from non-Gaussian 
signals of a non-primordial origin is a true challenge~\citep{Komatsu02,Komatsu03b}. 
It is well known that the theoretical CMB bi-spectrum is an optimal estimator 
for $f_{\rm \,NL}$~\citep{Babich,Komatsu2,Liguori,Ducout} and has been 
the most used tool for this purpose. 
However, alternative statistical estimators have been intensively studied in the last few years 
with diverse purposes such as to detect, to constrain, or simply to corroborate previous 
results claiming the presence of NGs in CMB data (see, e.g.,~\cite{Chiang,%
2007/Chiang,2013/modest}). 
One of the main reasons to employ other estimators is that one does not expect that any 
single statistical estimator can be sensitive to all possible forms and levels of NG.

In the last years, the CMB data from WMAP has undergone rigorous scrutiny for Gaussian 
deviations of any 
origin~\citep{Komatsu03b,Park,BTV1,BTV2,Chiang,2007/Chiang,McEwen,%
Wiaux,Bernui09,Cruz09,Pietrobon09,2009/rossi,Rossmanith,Vielva09,%
Pietrobon10,Raeth3,Bielewicz12,Gruppuso}. 
In particular, ef\/forts have been made to constrain the level of primordial NG 
in the successive WMAP releases~\citep{Komatsu03b,WMAP9b}, also using large-scale 
structure data~\citep{Verde00,Komatsu03a,Liguori,%
Yadav,Xia,Bartolo12,Chongchitnan,Takeuchi}. 
Because it is expected that primordial and non-primordial NGs are mixed in CMB data, 
the search for non-Gaussian signals in CMB maps includes galactic and extragalactic 
foregrounds, as well as contributions from systematic ef\/fects. 
The analyses of extragalactic foregrounds are mainly focused on secondary CMB 
anisotropies like the Sunyaev-Zel'dovich and lensing 
ef\/fects~\citep{Aghanim00,Cooray,Sadeh,Aghanim08,Aluri11,Cruz10,Munshi,%
Pace,Aluri12,Camila,Munshi13a}, along with point sources 
contamination~\citep{WMAP9a,PLA-XXVIII}. 
In this scenario, the application of various algorithms is recommended to discriminate 
between multiple forms of NG that could be present in CMB maps~\citep{Leach}, 
or at least to help to constraint their 
amplitudes~\citep{Casaponsa11a,Casaponsa11b,Ducout,%
Babich,BR09,Lew,2009/rossi,Vielva09,Noviello,Pietrobon09,Matsubara,%
NN,BR10,Pietrobon10,Raeth3,Rossmanith,%
Curto11,2011/rossi,Smith,BR12,Donzelli,Fergusson,Pratten,Zhao12}. 
%

Moreover, extensive works have been made by the WMAP team~\citep{WMAP7c,WMAP9a} 
and other groups~\citep{Naselsky,Chiang,2007/Chiang,Delabrouille,Cabella10,Saha}, to 
produce a foreground-reduced CMB map after separating the dif\/fuse galactic foregrounds. 
Recently, the {\it Planck collaboration} released three high resolution, almost full-sky, 
foreground-cleaned CMB maps~\cite{PLA-XII}. 
Each of these CMB maps was released together with realistic pixel's noise map and also 
a cut sky mask, outside which the CMB signal is considered statistically robust. 
The recent analyses performed by the {\it Planck collaboration} showed that, despite the agreement 
of the \textit{Planck} data sets with the Gaussian hypothesis, they do not exclude the presence 
of NGs of small intensity~\citep{PLA-XXIV}.

\begin{figure}
\includegraphics[width=.5\linewidth]{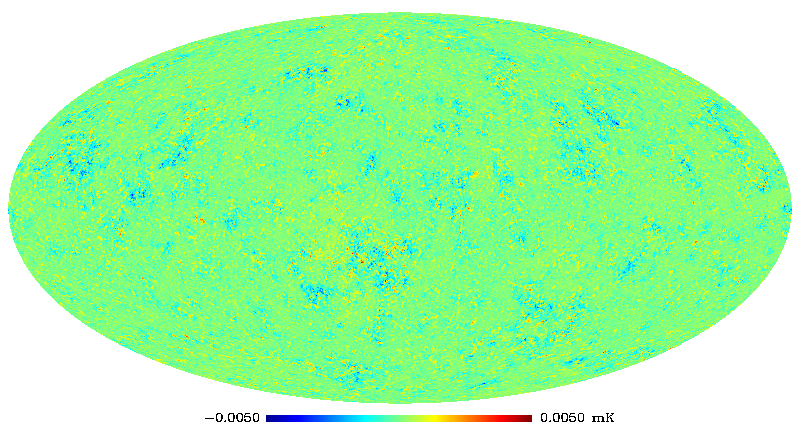}
\includegraphics[width=.5\linewidth]{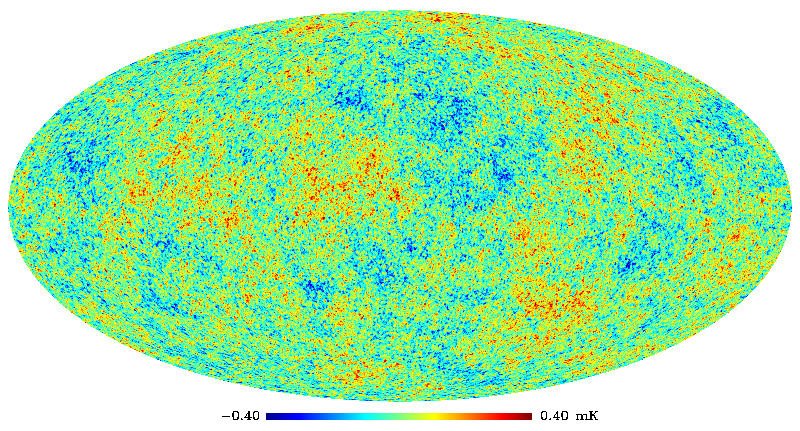}
\caption{\label{fig1} 
For illustration we show, in the upper panel, a simulated CMB map containing only a low 
level of local NG, namely 
$a_{\ell \, m}^{\mbox{\tiny upper}} = f_{\rm \,NL} \, a^{\mbox{\tiny NG}}_{\ell \, m}$ 
with $f_{\rm \,NL} = 38$. 
In fact, this level of NG is so low that to emphasize it we define the temperature limits to 
$[-0.005,0.005]$ mK in order to make the contamination visible. In the lower panel we have 
a combined map made up by a Gaussian CMB map plus the upper panel non-Gaussian map, 
i.e., $a_{\ell \, m}^{\mbox{\tiny lower}} = 
a^{\mbox{\tiny G}}_{\ell \, m} + a_{\ell \, m}^{\mbox{\tiny upper}}$, with the usual scale 
$[-0.4,0.4]$ mK. 
As can be realized, there is no a visible correlation of the NG present in both 
maps. Statistical estimators to reveal tiny NG are required.} 
\end{figure}

In this work we present an ef\/ficient statistical estimator, based on two powerful statistical 
tools, that proved to be sensitive enough to detect tiny primordial NGs of {\em local} type in 
Monte Carlo CMB maps. 
Our estimator combines the Minkowski functionals~\citep{1903/minkowski}, widely used to 
study the statistical CMB properties (see, e.g.,~\cite{Komatsu03a}), with an exhaustive 
analysis using a Neural Network, a 
computational tool for identifying patterns in a data set~\citep{1999/haykin}. 
For this purpose we produced sets of simulated CMB maps combining Gaussian and non-Gaussian maps containing dif\/ferent levels of {\em local} 
NG~\citep{Elsner} (for a review see, e.g.,~\cite{Bartolo04,Komatsu1,Komatsu2,Liguori}). 
We generate Monte Carlo CMB maps from the set of coef\/ficients 
$\{ a_{\ell \, m} \}$ which were produced according to the linear combination: 
$a_{\ell \, m} = a^{\mbox{\footnotesize G}}_{\ell \, m} + 
f_{\rm \,NL} 
\, a^{\mbox{\footnotesize NG}}_{\ell \, m}$
where $a^{\mbox{\footnotesize G}}_{\ell \, m}$ ($a^{\mbox{\footnotesize NG}}_{\ell \, m}$)
are the multipole expansion coef\/ficients  of the CMB Gaussian (non-Gaussian of local
type) map (see Figure \ref{fig1}).
The scalar coef\/ficient $f_{\rm \,NL}$ is the dimensionless parameter
commonly used to describe the leading-order of the NG. 
The Gaussian random sets $\{a^{\mbox{\footnotesize G}}_{\ell \, m}\}$, satisfying the 
angular power spectrum of the $\Lambda$CDM model, and the sets 
$\{a^{\mbox{\footnotesize NG}}_{\ell \, m}\}$ corresponding to a 
{\em local} type non-Gaussian CMB maps were taken from~\cite{Elsner}. 
According to the above linear combination of maps, one simulates CMB maps with 
an arbitrary level of NG of local type, defined by any real value of $f_{\rm \,NL}$.

One of the main challenges of an estimator is to discriminate between primordial and 
secondary (i.e., non-primordial) non-Gaussian signatures in a given CMB map. 
This is not an easy task. 
For instance, an analysis using Minkowski functionals partially confuses signals of these two 
different origins, introducing degeneracies in their diagnostic~\citep{PLA-XXIV}. 
Another important task for a statistical estimator is the effectiveness in the detection of tiny 
level of non-Gaussian contamination in CMB maps. 
Clearly, the suitability of a given statistical tool to perform these tasks deserves investigation. 

To analyze the potentialities of our estimator to reveal the presence of non-Gaussian signals 
in CMB data and to differentiate primary NG from secondary one, we produce several sets of 
thousands of simulated CMB maps, where one of these is a set of Gaussian CMB maps, and 
the other CMB maps are endowed with dif\/ferent amounts of primordial NG of {\em local} type. 
Thus, section~\ref{section2} contains the details of construction and operation of our 
combined statistical estimator. 
In section~\ref{section3} we apply our estimator to simulated non-Gaussian CMB maps, 
with several levels of contamination in dif\/ferent $f_{\rm \,NL}$ intervals in the range 
$[-10,80]$, as described  throughout the text, compatible with recent {\it Planck} constraints 
for local NG large angles.

Afterward, we perform robustness tests by applying our estimator to simulated maps using 
the masks released by the {\it Planck collaboration}, and adding also realistic inhomogeneous 
pixel noise maps (released with the corresponding masks). 
In section \ref{section4}, we analyze three foreground-reduced, almost full-sky, maps released 
by the {\em Planck collaboration}:  SMICA (Spectral Matching Independent Component 
Analysis), NILC (Needlet Internal Linear Combination), and SEVEM (Spectral Estimation 
Via Expectation Maximization). 
Finally, we present our conclusions and final remarks in section~\ref{section5}.


\section{The combined estimator: the Minkowski functionals and 
the Neural Networks tools} \label{section2}

The new estimator proposed to identify and quantify NGs in CMB data 
is a combination of two statistical tools: 
Minkowski Functionals (hereafter MF) and Artificial Neural Networks (NN). 
The MFs were introduced in cosmological studies as tridimensional statistics for the 
analysis of the CMB and the matter distribution in the universe (see, 
e.g.,~\cite{1987/vittorio_juszkiewicz, 1991/park_gott, 1994/mecke}), 
Describing the morphological properties of a given field, the MFs are being used as 
generic estimators of NG in the 2-dimensional CMB field. 
%
The use of NNs~\citep{1998/zhang_patuwo} as a statistical tool for CMB studies started to 
being applied more recently \citep{NN,2011/silva,2011/singal}. 
Specifically, regarding Gaussian deviations in CMB data, as far as we know, 
there is just one reference in the literature that make use of this tool, 
namely~\cite{Casaponsa11b}, 
where the authors combine the NN with wavelet analysis in order to estimate the non-linear 
dimensionless parameter $f_{\rm \,NL}$ using the constraints found in WMAP-9yr CMB data. 

In the next subsections we give a brief description of these tools and how we combine 
them to construct our estimator. 

\subsection{The Minkowski functionals applied to CMB maps} \label{section2.1}

All the morphological properties of a $d$-dimensional space can be described using 
$d+1$ MFs~\citep{1903/minkowski}. 
In the case of a CMB map, a 2-dimensional temperature field defined on the sphere, 
$\Delta T = \Delta T(\theta,\phi)$, with zero mean and variance $\sigma^2$, 
this tool provides a test of non-Gaussian features by assessing the properties of 
connected regions in the map. 
Given a sky path ${\cal P}$ of the pixelized CMB sphere ${\cal S}^2$, an {\em excursion set} 
of amplitude $\nu_t$ is defined as the set of pixels in ${\cal P}$ where the temperature field 
exceeds the threshold $\nu_t$, that is, it is the set of pixels with coordinates 
$(\theta,\phi) \in {\cal P}$ such that $\Delta T(\theta,\phi) / \sigma \equiv \nu > \nu_t$. 

In a two-dimensional case, for a region $R_i \subset {\cal S}^2$ with amplitude $\nu_t$ 
the partial MFs calculated just in $R_i$ are: $a_i$, the Area of the 
$R_i$ region, $l_i$, the Perimeter (contour length) of this Area, and $n_i$, the number of 
holes in this Area. 
The global MFs are obtained calculating these quantities for all the connected regions with 
height $\nu > \nu_t$. 
Then, the total Area $A(\nu)$, Perimeter $L(\nu)$ and Genus $G(\nu)$ are~\citep{1999/novikov,Komatsu03a,2006/naselsky_book,Ducout} 

\begin{eqnarray} \label{funcionais}
A(\nu) &=& \frac{1}{4 \pi} \int_{\Sigma} d\Omega = \sum a_i \, , \\
L(\nu) &=& \frac{1}{4 \pi} \frac{1}{4} \int_{\partial\Sigma} dl = \sum l_i \, , \\
G(\nu) &=& \frac{1}{4 \pi} \frac{1}{2 \pi} \int_{\partial\Sigma} \kappa dl = \sum g_i = 
N_{hot} - N_{cold} \, ,
\end{eqnarray}

\noindent
where $\Sigma$ is the set of regions with $\nu > \nu_t$, $\partial \Sigma$ is the boundary 
of $\Sigma$, and, $d \Omega$ and $dl$ are the elements of solid angle and line, 
respectively. 
In the genus definition, the quantity $\kappa$ is the geodesic curvature (for more details 
see, e.g.,~\cite{Ducout}). 
This last functional can also be calculated as the dif\/ference between the number of regions 
with $\nu > \nu_t$ (number of hot spots, $N_{hot}$) and regions with $\nu < \nu_t$ (number 
of cold spots, $N_{cold}$). As defined, the MF are calculated from a given threshold $\nu_t$.

The MFs are currently used to test the Gaussian nature of the CMB temperature 
f\/luctuations data~\citep{1999/novikov,Komatsu03a,Eriksen04,2006/naselsky_book,%
Curto08,Hikage12,2013/modest,Munshi13b}. 
As they are sensitive to weak and arbitrary non-Gaussian signals, e.g. small $f_{\rm \,NL}$ 
contaminations of different types, MFs are a complementary tool to optimal NG estimators 
based on the bispectrum calculations. 
Recently, the {\it Planck Collaboration} performed successful validation tests of the MF estimator 
with three sets of simulated CMB maps: Gaussian full-sky maps, full-sky non-Gaussian maps 
with noise, and non-Gaussian maps with noise and mask~\citep{PLA-XXIV}. 
Afterward, they used the MFs to quantify local NG in the foreground-cleaned {\it Planck} 
maps~\citep{PLA-XII}, where some instrumental effects and known non-Gaussian contributions, 
like lensing, were taken into account in the analyses using realistic lensed and unlensed 
simulations of {\it Planck} data~\citep{PLA-XXIV}. 
The constraints on local NG obtained are quite robust to Galactic residuals and are consistent to 
those from the bispectrum-based estimators. 
Moreover, these results, $f_{\rm \,NL} = 38 \pm 18$ for large angular scales, are basically equal 
to those obtained using WMAP-9yr data, that is $f_{\rm \,NL} = 37.2 \pm 19.9$~\citep{PLA-XXIV}.

At this point we would like to emphasize that here we use the MFs in a different way. 
As shown in section~\ref{test1}, we use the MFs as an intermediate tool: first we apply the MFs 
to a CMB map to discover that the {\it perimeter} is the MF with the better capability to capture 
the non-Gaussian signatures that, in a second step, are efficiently recognized by the NN. 
In this sense, we do not use MFs to quantify NG in the map. 
Instead we use it just to reveal the non-Gaussian imprints present in the maps, signatures that 
are then systematically recognized by the trained NN.


\subsection{The Neural Networks estimator} \label{section2.2}

The NN are computational techniques inspired in the neural structure of intelligent 
organisms (animal brains), which acquires knowledge through learning \citep{1999/haykin}. 
A NN is composed by a large number of processing units (also called artificial neuron or nodes), configured to perform a specific action, like pattern recognition and data classification. 

\begin{figure}
\centering
\includegraphics[scale=0.7]{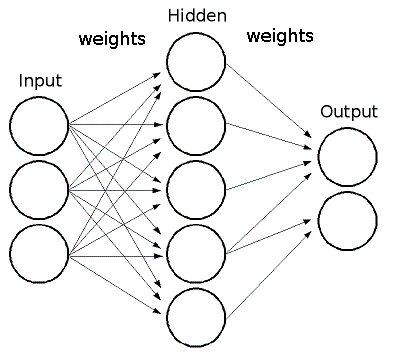}
\caption{Multilayer Perceptron.}
\label{NN}
\end{figure}

Aiming to emulate the behaviour of the biological brain, the simplest and most popular model 
for a NN for classification of patterns is the \textit{Perceptron} \citep{1943/mcculloch_pitts,1957/rosenblatt}, consisting of a single neuron. A generalization of this model is the 
\textit{Multilayer Perceptron} \citep{1986/rumelhart,1998/zhang_patuwo,1999/haykin}, consisting 
of a set of units (neurons) comprising each layer, from which the signal propagates through the 
NN. 
These NN are usually composed by an input layer, an output layer and one or more intermediate 
(or hidden) layers, as schematically depicted in Figure~\ref{NN}. 
The layers are interconnected through synaptic weights ($w_{ki}$) that relates the $i$-th input 
signal ($x_i$) to the $k$-th neuron, producing weighted inputs. 
Mathematically it is possible to represent a neuron $k$ by 

\begin{equation}
u_k = \sum_{i=1}^{n} w_{ki} ~.~ x_i + b_k,
\end{equation}

\noindent where $n$ is the number of input signals to the $k$-th neuron. $b_k$ is the weighted 
bias input, an external parameter of the artificial neuron $k$, which can be accounted for by 
adding a new synapse, with input signal fixed at $x_0 = + 1$ and weight 
$w_{k0} = b_k$~\citep{1999/haykin}.

The output signal of the $k$-th neuron ($y_k$) is generated through an activation function 
$\varphi_k$, which limits the amplitude of the output of a neuron 

\begin{equation}
y_k = \varphi_k(u_k) \, .
\end{equation}

\noindent The most commonly used activation function are non-linear functions, like sigmoid 
and hyperbolic tangent, that can simulate more precisely the neuron behaviour in order to 
better emulate a real NN. 

The suitable values for the {\em weights} and {\em bias} are achieved by using training (or 
learning) algorithms.  
The most popular training algorithm is the \textit{backpropagation} \citep{2000/basheer}, 
where the input signal feeds the first layer, propagating as input for the next layer, and so 
on until the last layer. 
At the end of this process a value for each neuron of the output layer, together with its corresponding error, is calculated. 
These values are returned to the input layer where the synaptic weights are 
recalculated initiating another \textit{iteration} of the training process. 
This procedure aims to determine a suitable value for the weights of the network, and is 
repeated until the error value drops below a given threshold value.
The architecture of the NN, like number of neurons, number of hidden layers and the 
specific training algorithm, can be defined according to the problem the user wants to solve. 
Here we used a backpropagation algorithm for the NN training with just one hidden layer, 
and the number of neurons was chosen according to the number of classes considered in 
the training: 80 neurons for 2 classes and 140 for 3 classes.

\subsection{The classifier estimator} \label{MF/NN}

The combined MF+NN estimator here presented shall be termed the {\em classifier} estimator. 
Regarding the MFs, they were calculated using the algorithm developed by~\cite{Ducout} 
and~\cite{2012/gay}. 
This code calculate four quantities, namely 
$V_0 = A(\nu)$, $V_1 = L(\nu)$, $V_2 = G(\nu)$, and $V_3 = N_{clusters}(\nu)$ 
the three usual MFs defined above plus an additional quantity called \textit{number of clusters}, 
$N_{clusters}(\nu)$, for $k = 3$. 
The latter quantity is the number of connected regions with height $\nu$ greater (or lower) 
than the threshold $\nu_t$ if it is positive (or negative), i.e., the number of hot (and cold) 
spots of the map. 

Consider a set of $m$ simulated CMB maps. 
For the $i$-th simulated map, with $i = 1, 2,..., m$, we compute the four MFs 
$\{V_k, \, k=0,1,2,3 \} \equiv (V_0,V_1,V_2,V_3)$ 
for $n$ dif\/ferent thresholds $\nu = \nu_{_{1}}, \nu_{_{2}}, ...\, \nu_n$, previously 
defined dividing the range $- \nu_{max}$ to $\nu_{max}$ in $n$ equal parts. 
Then, for the $i$-th map and for $k$-th MF we have not one element, but the vector 
\begin{eqnarray} \label{def_v}
\,\, v_k^i \equiv ( V_k(\nu_{_{1}}),V_k(\nu_{_{2}}), ... V_k(\nu_n) )|_{\mbox{for the $i$-th map}} \, .
\end{eqnarray}
In this work the values chosen for such variables are 
$\{\nu_{max},n\} = \{3.5,26\}$~\citep{Gott,Ducout}.

Once calculated the MFs vectors we define the training data set, $T\{x_i,y_i\}$, for the NN, 
where $x_i$ is called input data and $y_i$ the output data, for $i = 1, 2, ..., m$, and $m$ is the 
number of simulated CMB maps. 
This training set configuration is necessary in order to allow the network to associate
a certain kind of input pattern to a specific output.
We set the $i-th$ input vector as the MF vector of the $i-th$ 
simulated map, $x_i = v_k^i$, for just the $k$-th MF. 
The output vector, $y_i$, is defined according to the number of classes $N_{class}$ of the 
input data. 
For example, if we use $N$ $classes$ of maps, 
corresponding to ensembles with dif\/ferent non-Gaussian levels, we have $N_{class} = N$. 
Then, our $m$ output vectors have $N$ elements, $y_i = (1,0,...,0)$ for $class = 1$, 
$y_i = (0,1,...,0)$ for $class = 2$, and so on, until $y_i = (0,0,...,1)$ for $class = N$. 

After this training process, the NN should be able to identify the same pattern in a dif\/ferent 
set of input vectors, e.g. the test data set $\{x_j,y_j\}$, for $j= 1, 2, ..., l$. 
That is, providing to the already trained NN a data set with $l$ input vectors $\{x_j = v_k^j\}$, 
it returns $l$ output vectors $\{y_j\}$ with a specified classification. 
It means that the NN classifies each CMB map --whose MF information is contained in 
$\{x_j\}$-- according to the class it belongs, in other words, the NN {\em classifier} estimator 
informs its non-Gaussian level ($f_{\rm \,NL}$). 
After the calculation of the output vectors $\{y_j\}$ for a given set of simulated CMB maps, we define a way to quantify the ef\/ficiency of the NN through the counting of the number of successes (or hits) of the NN in classifying each one.
We find out two ways of counting the \textit{hits} of the NN: \\
\textit{Mode-1:} \,According to the limit value of 0.5 for each element of $y_j$ vector. 
In this case, e.g. for $N_{class} = N$, a vector with elements like $y_j = ({\bf >0.5},<0.5,...,<0.5)$ 
indicates the $class = 1$, a vector like $y_j = (<0.5, {\bf >0.5},..., <0.5)$ indicates 
$class \!=\! 2$, and so on, until $y_j = (<0.5, <0.5,..., {\bf >0.5})$ that indicates $class = N$. \\
\textit{Mode-2:} \,Considering that the largest of the three elements of an output vector, $y_j$, 
indicate the class to which it belongs. 
For example, again for $N_{class} = N$, if the first element of the vector is the largest, this 
output vector indicates $class = 1$, if the second is the largest, it indicates $class = 2$, 
and so on, until $class = N$. \\
Thus, we quantify the hits of our NN by classifying each analyzed map according to its 
$class$, in other words according to its $f_{\rm \,NL}$ value.

\section{Applying our estimator to non-Gaussian CMB maps} \label{section3}

\subsection{The Monte Carlo CMB maps} \label{section3.1} 

An optimal network training requires a large amount of simulated data. 
We produce simulated maps, termed Monte Carlo (MC) CMB maps, 
cut off at $\ell_{max} = 500$, with $N_{\mbox{\small side}} = 512$, 
using the HEALPix (Hierarchical Equal Area iso-Latitude Pixelization) 
pixelization grid~\citep{Gorski}, using the publicly available set of 1000 linear, 
$\{ a^{\mbox{\footnotesize G}}_{\ell \, m} \}$, and 1000 non-linear (corresponding to NG 
of {\em local} type), $\{ a^{\mbox{\footnotesize NG}}_{\ell \, m} \}$, spherical harmonics 
coef\/ficients, produced 
by~\cite{Elsner}\footnote{http://planck.mpa-garching.mpg.de/cmb/fnl-simulations/}. 
We combine them as follows 

\begin{equation}
\, a_{\ell \, m} = a^{\mbox{\footnotesize G}}_{\ell \, m} 
+ f_{\rm \,NL} \, a^{\mbox{\footnotesize NG}}_{\ell \, m} \, ,
\end{equation}

\noindent 
and normalize the maps using a power spectrum generated by the 
CAMB\footnote{http://lambda.gsfc.nasa.gov/toolbox/tb\_camb\_form.cfm} online tool 
using the {\it Planck} concordance cosmological model~\citep{PLA-I}. 

To test the ef\/ficiency of our estimator we use MC CMB maps containing 
the degree of contamination defined by the recent constraints found by 
{\it Planck}~\citep{PLA-XXIV}, that is, $f_{\rm \,NL} = 38 \,\pm\, 18$ at 68.3\% confidence 
level. 
For this, we test our estimator with sets of MCs endowed with levels 
of contamination in the ranges: $f_{\rm \,NL} \,=\, [-10,10], [10,30], \,[28,48], [30,50], [40,60], \,[60,80]$, in order to have normal distributions with mean 0, 20, 38, 40, 50 and 70, respectively.

\subsection{Quantitative analysis of NG in CMB maps} \label{section3.2}
\subsubsection{Preliminary tests: sensitivity of the classifier estimator} \label{test1}

We begin the tests using just two $classes$, corresponding to MC CMB maps constructed 
using $f_{\rm \,NL} = [-10,10]$ ($class = 1$) and $f_{\rm \,NL} = [40,60]$ ($class = 2$). 
We started using $m = 1 000$ Area vectors ($v_0^i$, for $i$ = 1, ..., 1000) for training the NN, 500 of each class. 
The output vectors were defined as, $y_i = (1,0)$ for $class = 1$ and $y_i = (0,1)$ for 
$class = 2$. 
After trained, the NN was applied to a test data set, of the same size as the training one, $l=m=1000$ Area vectors, with again 500 of each class.
This procedure of training and test of NNs was repeated for each one of the three other kinds of MFs ($k = 1, 2$, and $3$).

Analyzing the output returned by the NN when applied to test data set, for each sort of MF, we had the results summarized in Table \ref{tab:results_first_tests}. 
The last two columns are the NN Hits, indicating the percentage of correct indication (right pattern recognition) according to both Modes 1 and 2, and the corresponding Mean Square Error (MSE), that measure the NN's performance comparing the returned values ($y_j^{ret}$) to those expected 
($y_j^{exp}$), defined as

\begin{equation}
MSE = \frac{1}{l} \sum_{j=1}^l (y_j^{exp} - y_j^{ret})^2 \, ,
\end{equation}
 
\noindent for $j=1, 2, ..., l$, where $l$ is the number of MF vectors in the test data set. 
These preliminary tests allow the comparison of sensitivities for each MF when used together 
with the NN {\em classifier} estimator. 
From these results it is possible to infer that the sensitivity of MFs, from the best to worst, is 
$\mbox{Perimeter} \, \gtrsim \, \mbox{Genus} \, \gtrsim \, N_{clusters} \, \gg \, \mbox{Area}$, 
very similar to what was obtained by~\cite{Ducout}. 

Using $m=2 000$ MF vectors 
to train our NN, each set of $1 000$ corresponding to the same classes 1 and 2, it was 
possible to check their sensitivity when increasing the size of the training data set. 
The results, also presented in Table \ref{tab:results_first_tests}, show a small improvement 
relatively to the test using only $1 000$ MF input vectors. 
In the next subsection we show the results from a second test, making clear how this feature 
improves the accuracy of our results. 

\begin{table*}  
\centering 
\caption{Results of the sensitivity tests.} {\footnotesize
\begin{tabular}{c c c c c c c}
\hline \hline
f$_{\mbox{\sc\footnotesize nl}}$ & m & l & $N_{class}$ & MF & Hits -\textit{ Mode-1/Mode-2} $^a$ (\%) & MSE $^b$ \\
\hline
                 &      &     &             & Area           & 59.4 / 63.7       & 0.240 \\ 
\cline{5-7}
[-10,10], [40,60]& 1000 & 1000 &       2     & Perimeter      & 85.5 / 87.1       & 0.106 \\ 
\cline{5-7}
                 &      &     &             & Genus          & 78.5 / 81.0       & 0.146 \\ 
\cline{5-7}
                 &      &     &             & $N_{clusters}$ & 74.5 / 81.5       & 0.146 \\ 
\hline
                 &      &     &             & Area            & 62.7 / 64.5       & 0.226 \\ 
\cline{5-7}
[-10,10], [40,60]& 2000 & 1000 &      2      & Perimeter       & 96.4 / 96.5      & 0.049 \\ 
\cline{5-7}
          &      &     &             & Genus           & 96.7 / 96.9      & 0.049 \\ 
\cline{5-7}
                 &      &     &             & $N_{clusters}$  & 95.7 / 95.9      & 0.062 \\ 
\hline
\hline
\multicolumn{7}{p{8cm}}{$^a$The two ways to calculate the hits of NN (see sect. \ref{MF/NN}). \newline $^b$Mean Square Error (see text for details).} 
\end{tabular}} \label{tab:results_first_tests}
\end{table*}

\subsubsection{Testing the estimator with lower non-Gaussian contamination level and 
greater number of classes} \label{test2}

We find out that the minimal separation between two neighbor $classes$, in order to 
guarantee a successful classification of $\sim 95\%$ of tested maps, is 
$\Delta f_{NL} \sim 35$ which gives the 2-$\sigma$ error of our method. 

As recently obtained from {\it Planck} data, the current limit for the non-linear 
parameter is $f_{\rm \,NL} = 38 \pm 18$, at 68\% confidence level. 
Our specific aim here is to test an estimator sensitive enough to identify the presence 
of weak levels of NG within this constraint. 
Therefore, taking into account the error of our method and the Planck constraint, 
we shall test our {\em classifier} estimator using three classes, $N_{class} = 3$, of 
simulated maps: 
one of nearly Gaussian CMB maps, $f_{\rm \,NL} = [-10,10]$ ($class = 1$, 
$\langle f_{\rm \,NL} \rangle = 0$), and the other made with two dif\/ferent levels of local 
NG: with $f_{\rm \,NL} = [28,48]$ ($class = 2$, $\langle f_{\rm \,NL} \rangle = 38$) and 
$f_{\rm \,NL} = [60,80]$ ($class = 3$, $\langle f_{\rm \,NL} \rangle = 70$), 
all three ensembles having the variance equal to 7.5.

According to our results obtained in the last sub-section the size of the training set appears 
to be a determinant factor in obtaining a good classifier, besides the fact that, between the 
four MFs, the {\em Perimeter} appears to be the most sensitive to reveal the non-Gaussian signal. 
Then, we started applying our estimator in $m = 3000$ Perimeter vectors, $l = 1000$ of each class. 
The same was repeated for $m= 4500$, $6000$, $7500$, $9000$, $10500$, $12000$, 
$13500$, $15000$, $16500$, $18000$ and $19500$, $m/3$ vectors of each class. 
For all applications we carried out, it was used a set of $l/3=500$ input Perimeter vectors in order to 
test the trained NN. 

\begin{figure}
\centering
\includegraphics[scale=0.5]{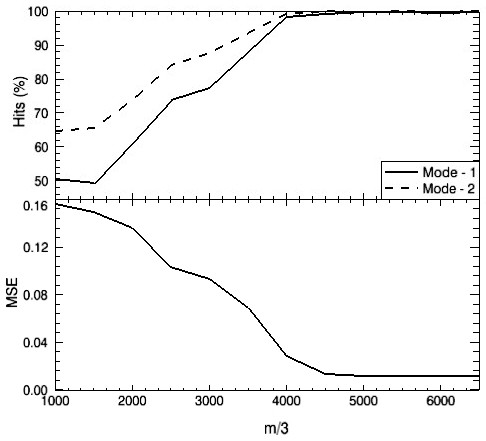} 
\caption{Average Hits (above) and Mean Square Error (bellow) in estimating the $f_{\rm \,NL}$ range using a NN previously trained with $m$ Perimeter vectors (Gaussian and non-Gaussian MC CMB maps).}
\label{dif_n}
\end{figure}

Figure~\ref{dif_n} shows the average number of hits resulting of each NN training, 
considering the two ways to count it, that is, Mode-1 and Mode-2. 
With this graph it is possible to verify that the larger the training data set the better trained 
will be the NN and more precise the classification. 
However, we can also identify a point ($m/3 \sim 4500$) where the hits do not improve 
significantly even further increasing the training set. 
Hence, we stopped our test with $m=19500$, with an average hits of 99.6\%, according Mode-1, 
and 99.9\%, according to Mode-2. The hits for each one of the classes 1, 2, 
and 3 are, respectively: 99.6\%, 99.8\% and 99.6\%, according to Mode-1, and 100\%, 
99.8\% and 100\%, according to Mode-2.

\subsection{Robustness analyses using Planck data} \label{Planck maps}%
\label{section3.2}

\subsubsection{Additional products from foreground-cleaned {\it Planck} maps}

The major challenge in making foreground-reduced CMB maps is to clearly understand  
all the possible sources of contaminations contributing to the set of dif\/ferent frequency maps 
observed in a given CMB survey. 
This is necessary to estimate their ef\/fects on the data in order to subtract the foreground 
components from the raw maps and to leave the largest possible amount of primordial cosmological signal. 
The {\em Planck collaboration} uses the suitable framework for this task, that is, the 
{\em component separation technique}~\citep{PLA-XII}, which aims to identify the sources 
of contaminating emissions that contribute dissimilarly to the set of nine frequency bands, 
from 30 to 857 GHz, observed by {\it Planck}~\citep{PLA-I}. 
In fact, four distinct component separation algorithms~\citep{PLA-XII} were employed by the 
{\em Planck collaboration} to identify and remove the signal coming from the various 
contaminants. 
At large angles, galactic foregrounds are dominant, including dif\/fuse emission 
(synchrotron, bremsstrahlung, thermal dust, and anomalous microwave emission) and line 
emission from carbon monoxide. 
On the other hand, extragalactic foregrounds dominate at small scales and can be classified into two 
categories: discrete detectable compact sources and the collective signal from unresolved 
sources (mainly radio and infrared galaxies, but also from clusters of galaxies via the 
Sunyaev-Zel'dovich ef\/fect). 
It is worth noticing that the {\em Planck} team just masked the bright cosmic objects  to eliminate 
extragalactic foregrounds.

As a result, the {\it Planck collaboration} has released three high resolution, almost 
full-sky foreground-cleaned CMB 
maps\footnote{http://pla.esac.esa.int/pla/aio/planckProducts.html} ~\citep{PLA-XII}, termed SMICA (Spectral Matching Independent Component Analysis)~\citep{Cardoso08}, NILC (Needlet Internal 
Linear Combination)~\citep{Delabrouille}, and SEVEM (Spectral Estimation Via Expectation 
Maximization)~\citep{Fernandez-Cobos12}. 
These high resolution maps have $N_{side} = 2\,048$ and ef\/fective beam 
{\sc fwhm}$ = 5$ arcmin. 
The performance of the foreground-cleaning algorithms has been carefully investigated with 
simulated maps. 
In particular, each component separation algorithm processes dif\/ferently the multi-contaminated 
sky regions obtaining, as a consequence, dif\/ferent final foreground-cleaned regions for each 
procedure. 
These regions are defined by the so-called Component Separation Confidence masks 
(brief\/ly termed CS-masks), which have $f_{sky} = 0.89, 0.93, 0.76$, for the SMICA, NILC, 
and SEVEM maps, respectively. 
Each CMB map was released with its own CS-mask, outside which the corresponding 
CMB signal is considered statistically robust. 
Regarding the masks, it is worth mentioning that the \textit{Planck} team produced the U73 mask, which 
is the union of the above mentioned masks, it is more restrictive since it has 
$f_{sky} = 0.73$ and is used for evaluation purposes~\citep{PLA-XII}. 
In addition, the mapmaking of these foreground-cleaned maps produced an estimative of 
the noise (mainly due to the non-uniform strategy for scanning the CMB sky) in each 
pixel's map, information that was also released jointly to its corresponding CMB map. 

The availability of these CS-masks and pixel noise maps allows to test the NG statistical 
estimators in non ideal situations, aiming to mimic some {\it Planck} data features. 
This is done analyzing simulated maps which include both ef\/fects: masks and noise, 
in addition to {\it local} NG. 
Nevertheless, this is a partial approach, because other contaminating contents, which are 
not taken into account in the simulations, could be present in the data. 
The consequences of this ignorance about unknown NGs in data maps will be discussed 
below. 

\subsubsection{Noise contamination and mask inf\/luence to the classifier estimator} 
\label{section3.3}

The aim of the following tests is to verify the performance of our estimator in more realistic 
situations, using for this dif\/ferent masks and realistic inhomogeneous pixel noise 
released by {\it Planck}. 
Moreover, in all these tests we take into account the {\it Planck} ef\/fective beam, 
{\sc fwhm}$ = 5$ arcmin, in the construction of the MC CMB maps. 

Since the previous tests (Figure \ref{dif_n}) showed that for $m/3 > 4500$ Perimeter-MF 
vectors the number of hits do not improve significantly, we decided to perform all the next 
tests using $m/3 = 5000$ MC CMB maps. For the tests of all the following trained NN it 
was used $l/3 = 500$ Perimeter-MF vectors. 

We divide the analyses in two steps: the first one serves to verify the impact of using masks, 
with dif\/ferent $f_{sky}$ factors, and the other one to check the estimator performance when 
the maps are contaminated by inhomogeneous noise, beside using masks.

\vspace{0.5cm}
\textit{\textbf{1. Sky cuts: the Planck masks}} 

The firsts robustness tests of the NN {\em classifier} estimator consists on the analysis of 
MC CMB maps with two sky cuts, given by two masks released by {\it Planck} Satellite: 
1) the CS-mask of the SMICA map, hereafter called SMICA-mask; and 2) the U73 mask. 
Therefore, the test of the masks were performed applying the estimator to two sets of maps, each one composed by a dif\/ferent sky fraction. Both tests used three classes of maps, composed by a non-Gaussian signal such as $f_{\rm \,NL} = [-10,10]$ ($class = 1$), $[28,48]$ ($class = 2$) and $[60,80]$ ($class = 3$). 
The average hits resulting from these tests, that is the rate of successfully detection cases, 
are shown in Table~\ref{tab:results_robustness_tests}.

Comparing the values presented in the second and third columns of 
Table~\ref{tab:results_robustness_tests} with the values of Table~\ref{tab:results_first_tests} 
we conclude that the use of masks does not inf\/luence the ef\/ficiency of the NN {\em classifier} 
estimator, and the average hits remains larger than 99\%. 

\vspace{0.5cm}
\textbf{\textit{2. Inhomogeneous noise and masks}}

A second set of robustness analyses consists on the application 
of our estimator to MC CMB maps contaminated with real non-Gaussian inhomogeneous noise 
from {\it Planck} data. 
In order to exhaustively test our combined estimator, various combinations of noise maps and 
dif\/ferent masks were applied to the MC CMB maps before calculating the input Perimeter 
vectors. 
These pixel noise maps are those corresponding to the SMICA, NILC, and SEVEM 
{\it Planck} maps, and the masks are the same used in the previous tests. 
Again it was used three \textit{classes} of synthetic maps, constructed using the same 
ranges of $f_{\rm \,NL}$ values used in previous analysis, that are: $f_{\rm \,NL} = [-10,10]$, 
$[28,48]$ and $[60,80]$. 
Then, it was performed a total of four tests of the estimator, in four dif\/ferent sets of synthetic 
maps, that are: two sets of maps contaminated by SMICA noise, using 1) the U73 mask and 
2) their own CS-mask, to analyze the inf\/luence of the two kinds of masks in the presence 
of the same noise, and two sets of maps contaminated with 3) SEVEM and 4) NILC noises, 
but using the same mask, U73, testing the impact of these two kinds of noise even with a 
rigorous sky cut. 

In addition, we still tested the ef\/ficiency of our estimator when applied to MC CMB maps 
whose primordial non-Gaussian signal corresponds to contiguous $f_{\rm \,NL}$ ranges, 
namely, $[-10,10]$,  $[10,30]$ and $[30,50]$. 
This test was carried out just in the case of using the SMICA noise as contaminant and the 
U73 mask. 

The hits numbers for each test, as well as the corresponding MSE
values, are presented in Table~\ref{tab:results_robustness_tests}. 
Again, there is no impact in using dif\/ferent masks, but the inclusion of inhomogeneous noise 
makes the number of hits become slightly lower. 
These results allow us to conclude that our combined estimator is, basically, equally ef\/ficient 
in both situations analyzed, namely applying a mask but not noise and using a mask plus a 
pixel noise map. 

The last point to discuss are the results obtained in Table~\ref{tab:results_robustness_tests} 
when using input maps endowed with primordial non-Gaussian signal in contiguous ranges of 
$f_{\rm \,NL}$ values, namely $[-10,10]$, $[10,30]$ and $[30,50]$. 
In this case the hit's numbers are even lower, with higher MSE value. 
This fact evidences the dif\/ficulty of the NN to identify non-Gaussian patterns when trained 
with sets of MCs having contaminations whose intensities $f_{\rm \,NL}$ belong to three 
adjacent intervals. 

%

\begin{table}
\centering
\caption{Results of the robustness analysis.} {\footnotesize
\begin{tabular}{c c c c c}
\hline \hline
$f_{\rm \,NL}{^a}$            & NN reference & MC CMB map                 & Hits$^b$ (\%)   & MSE  \\
\hline
                              & NN$^{nc}$-1  & noiseless   + SMICA-mask   & 99.4 / 99.7     & 0.01 \\
                              & NN$^{nc}$-2  & noiseless   + U73          & 99.6 / 99.7     & 0.01 \\
 $[-10,10], [28,48], [60,80]$ & NN$^{nc}$-3  & SMICA noise + SMICA-mask   & 98.4 / 99.1     & 0.03 \\
					          & NN$^{nc}$-4  & SMICA noise + U73          & 97.3 / 98.5     & 0.02 \\
				              & NN$^{nc}$-5  & SEVEM noise + U73          & 98.1 / 98.9     & 0.03 \\
				              & NN$^{nc}$-6  & NILC noise  + U73          & 98.6 / 99.3     & 0.03 \\
\hline
$[-10,10],[10,30],[30,50]$    & NN$^{c}$-7   & SMICA noise + U73          & 96.5 / 97.2     & 0.05 \\
\hline
\hline
\multicolumn{5}{p{150mm}}{$^a$ Contiguous ($c$) and not contiguous 
(${nc}$) intervals of the parameter $f_{\rm \,NL}$ in the MC simulations. 
\newline $^b$ Hits corresponding to \textit{Mode-1/Mode-2}.}
\end{tabular}} \label{tab:results_robustness_tests}
\end{table}

\section{Analysis of Planck CMB maps} \label{section4}

The purpose, and natural consequence, of the development of a new statistical tool, after exhaustive tests in synthetic data, is its application to real data. 
As discussed before, the robustness tests presented here were done using {\it Planck} data, 
like real pixel noise maps and cut-sky masks, 
because our aim is to know the impact of these effects on our estimator for a posterior 
application of our trained NN to corresponding CMB data.
For each foreground-cleaned {\it Planck} CMB map we used the suitably trained NN, 
that is, that one trained using MCs CMB maps to which we added pixel noise contamination 
and to which we apply a sky-cut mask. 
Thus, the analysis of the three {\it Planck} CMB maps were performed according to the 
following correspondence: 
\begin{eqnarray} 
\mbox{NN}^{nc}\mbox{-3} & \Longrightarrow & \mbox{SMICA map + SMICA-mask} \label{correspondences1} \\
\mbox{NN}^{nc}\mbox{-4} & \Longrightarrow & \mbox{SMICA map + U73} \label{correspondences2} \\
\mbox{NN}^{nc}\mbox{-5} & \Longrightarrow & \mbox{NILC map + U73} \label{correspondences3} \\
\mbox{NN}^{nc}\mbox{-6} & \Longrightarrow & \mbox{SEVEM map + U73} \label{correspondences4} \\
\mbox{NN}^{c}\mbox{-7} \,\,\,& \Longrightarrow & \mbox{SMICA map + U73} \label{correspondences5}
\end{eqnarray}

\noindent where the indices 3-7 indicate the trained-NN used in each case (according to the  input maps used for each one), and $c$ ($nc$) refers to the contiguous (not contiguous) 
intervals employed in the analyses, as specified in Table~\ref{tab:results_robustness_tests}. 

\begin{figure}
\centering
\begin{minipage}[b]{0.4\linewidth}
\includegraphics[width=\linewidth]{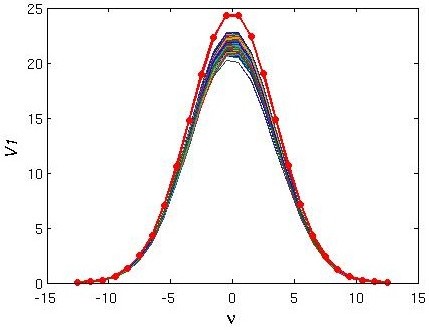}
\end{minipage}
\begin{minipage}[b]{0.4\linewidth}
\includegraphics[width=\linewidth]{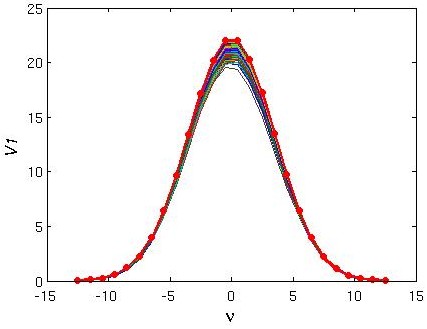}
\end{minipage}
\caption{Plot of the Perimeter-MF calculated from the MC CMB maps (multicoloured lines) and from SMICA {\it Planck} map (red thick line). Left: Perimeter vectors calculated from the original maps after applying the U73 mask. Right: The same as in the left, but all the maps is smoothed with {\sc fwhm}$ = 3$ arcmin before calculate the Perimeter.} 
 \label{perimeter}
\end{figure}

As discussed before, the expected output of a NN is a vector whose elements are 
$\cal{O}$(0) or $\cal{O}$(1), where $\cal{O}$(0) ($\cal{O}$(1)) means that the numerical 
value is very near to zero (to one). 
That is, in our case, using $N_{class} = 3$, 
$y_{\mbox{\tiny Planck map}} =$ ($\cal{O}$(1), $\cal{O}$(0), $\cal{O}$(0)), or 
$y_{\mbox{\tiny Planck map}} =$ ($\cal{O}$(0), $\cal{O}$(1), $\cal{O}$(0)), or 
$y_{\mbox{\tiny Planck map}} =$ ($\cal{O}$(0), $\cal{O}$(0), $\cal{O}$(1)), indicating the 
class of a given map. 
However, the NN returned $y$ vectors with elements dif\/ferent from those 
expected, presenting values of order $\cal{O}$(10), or even higher. 
These kind of discrepancy occurs when using training and test data sets with very dif\/ferent characteristics. 
Then the NN response is totally comprehensive when comparing a Perimeter vector from 
{\it Planck} CMB map to that ones obtained from the MC CMB maps, which are visually 
dif\/ferent, as shown in Figure~\ref{perimeter}. 

It is worth to emphasize that the MFs are sensitive to the resolution scale of the 
temperature map in analysis~\cite{hikage/2008}. 
In this sense, a possible explanation for the difference in the Perimeter's amplitudes 
observed between synthetic and Planck maps (left panel of Figure~\ref{perimeter}) is that, 
even performing MC simulations using the \textit{Planck} effective beam of 
{\sc fwhm}$ = 5$ arc-min, the smoothing scales of these maps are still not fully compatible. 
Aware of this problem, and attempting to solve it, we analyze the ef\/fect of using a 
smoothing tool in all the MC CMB maps (training and test data sets), as well as in the 
{\it Planck} maps, before the calculation of the MFs. 
It was verified that this procedure makes the characteristics of these maps a bit more 
consistent, obtaining more compatibles Perimeter's vectors (right panel of 
Figure \ref{perimeter}). 
The Gaussian beam used for smoothing all the maps was {\sc fwhm}$ = 3$ arcmin. 
This value was chosen as a compromise: 
it is large enough in such a way that synthetic and real maps became more consistent, 
and it is small enough to not decrease the amplitude of the non-Gaussian signal, under 
the condition that does not modify the angular power spectrum of the maps. 
It is worth to point out that this procedure does not inf\/luence the results of applying 
the NN to the test data set, still obtaining hits numbers as those presented in 
Table~\ref{tab:results_robustness_tests} for all the cases. 
Moreover, when applying the NN, trained using Perimeter vectors from these slightly 
smoothed maps, to those obtained from the smoothed {\it Planck} CMB maps, we 
verify a significant improvement of the output vectors. Following the same correspondences 
in~\ref{correspondences1}-\ref{correspondences5}, the vectors obtained were, respectively:
\begin{eqnarray} 
 y_{\tiny \mbox{NN}^{nc}\mbox{-3}} &=& (-0.45,0.99,0.46) \label{output1}\\
 y_{\tiny \mbox{NN}^{nc}\mbox{-4}} &=& (-0.11,0.75,0.35) \label{output2}\\
 y_{\tiny \mbox{NN}^{nc}\mbox{-5}} &=& (-0.25,1.97,-0.72) \label{output3}\\
 y_{\tiny \mbox{NN}^{nc}\mbox{-6}} &=& (-0.37,-3.61,4.98) \label{output4}\\
 y_{\tiny \mbox{NN}^{c}\mbox{-7}} \,\;&=& (0.55,-2.11,2.56) \label{output5}
\end{eqnarray} 
As observed, there is an improvement in the consistency of the output vectors $y$, but it is 
not enough to give us a reliable result because the elements are still quite dif\/ferent from 
those expected. 

We interpret these results in the sense that, under these \textit{circumstances}, the NN is not able to recognize the full pattern of NG appearing in the {\it Planck} maps. These \textit{circumstances} means that even including noise, using masks and running a smoothing procedure, in such a way to let the MC CMB maps more similar to {\it Planck} CMB maps, the synthetic maps will not mimic them well enough, since we do not have a total knowledge about the full contents of these maps. 
Moreover, and complementing this idea, 
the imprecision of the NN could be indicating that the NGs present in the {\it Planck} maps 
are not only of local type. 
In this case, since the simulated data do not have contributions apart from the local 
non-Gaussianity, the estimator shows that the synthetic and \textit{Planck} non-Gaussianities 
contents are, indeed, different.

Nevertheless, even when the results from the application of our NNs into 
{\it Planck} CMB maps reveal certain imprecision of the NN, one can still verify that the output 
vectors indicate a level of non-Gaussian contamination that agree with the recent results obtained with WMAP-9yr and {\it Planck} data. Analysing these output vectors in the sense that its higher element denote the class of the corresponding map (as in {\it Mode-2}), that is, the range of 
$f_{\rm \,NL}$ values that the NN indicates for the non-Gaussian contamination, we get the 
ranges: $f_{\rm \,NL} = [28,48]$ for the output vectors of relations (\ref{output1}) to 
(\ref{output4}), and  $[30,50]$ for the output vector of the relation (\ref{output5}). 
It is worth mentioning that in some approaches it is not possible to exhibit the results in 
the form: average value (or most probable value) $\pm$ 
one standard deviation, i.e., $\overline{\mbox{X}} \pm \Delta \mbox{X}$. 
In such a case, alternative ways are employed to measure the efficiency 
of a method (as e.g. in \cite{2009/rossi,Rossmanith,Raeth3,2013/modest}). 
For this reason, here we use the MSE value, which is the standard criterion often used to 
measure the performance of a NN (as well as RMSE, the squared root of MSE), and, therefore, 
an appropriate error measurement to evaluate our estimator (see, 
e.g.,~\cite{1998/zhang_patuwo,2000/zhang,claudio,2013/Casaponsa}). 


\section{Conclusions} \label{section5}

The standard inf\/lationary model predicts a Gaussian distribution of CMB anisotropy, while other theories predict the existence of deviations from the initial Gaussian condition. 
The detection, or not, of non-Gaussian (NG) signatures on CMB data can help to exclude many inf\/lationary models and discriminate between dif\/ferent mechanisms for generating cosmological perturbations. 
Therefore, the search for non-Gaussian signals in new CMB data sets has become the target 
of several working groups, analyzing so far mainly the WMAP 
data~\citep{Komatsu03b,Park,Chiang,2007/Chiang,McEwen,%
Wiaux,Bernui09,Cruz09,Pietrobon09,2009/rossi,Rossmanith,Vielva09,%
Pietrobon10,Raeth3,Bielewicz12,Gruppuso,Cabella10}. 
Currently, with the recent release of {\it Planck} data, analyses following dif\/ferent approaches 
are being done to confirm independently the constraints on primordial NG found by the 
{\it Planck team}, taking into account that many estimators are needed to identify diverse 
types of NG, their intensities, and corresponding angular scales 
dependences~\citep{PLA-XXIV,Komatsu02,Ducout,2013/modest,Raeth3}.

We presented here an estimator developed to improve the determination of constraints on the primordial NG, gathering a tool already widely used to study non-Gaussian signal, the Minkowski Functionals, and another one complementary and very new in this context, the Artificial Neural Networks. The joint use of these tools provide a classifier estimator which performs a comparison analysis, as described throughout the text. The estimator was applied to a large amount of MC CMB simulations divided in classes according to the non-Gaussian level of contamination, that is, the $f_{\rm \,NL}$ value. One conclusion was that the higher the number of maps (MF vectors) used as input the better the performance of the estimator, but with a limit in which the results stop to improve (Figure \ref{dif_n}). We also tested the joint use of each of the four kinds of MF with the NN, and from the results of each case it was possible to verify that the Perimeter-MF is the most sensitive to non-Gaussian signal, providing a pattern more easily recognized by the NN 
(see Table \ref{tab:results_first_tests}).

A large number of tests were performed in order to evaluated the performance of the estimator 
in dif\/ferent configurations, and to verify what can inf\/luence the results. We notice that the use 
of masks have no impact, and the inclusion of inhomogeneous noise does not af\/fect significantly 
as well. 
Moreover, the performance of the estimator can be inf\/luenced by the number of classes and 
the degree of NG.  When using just two classes, corresponding to MC CMB maps constructed 
using $f_{\rm \,NL}$ values more widely spaced, it is not necessary to use a very large number 
of neurons, whereas using three classes require a larger number of them in order to train 
the NN. 

As a matter of fact, the capability to recognize primary and secondary CMB NGs, besides 
having a good performance in capturing weak NG contaminations in the data are essential 
features in a statistical estimator. 
For this, we analyzed the capacity of our combined estimator to distinguish between 
primary and secondary NGs. 
Training the Neural Network with two sets, one of nearly Gaussian CMB maps 
($|f_{\rm \,NL}| \le 10$) endowed with realistic inhomogeneous {\it Planck} noise 
(i.e., secondary non-Gaussianity) and the other of CMB maps with weak primordial NG 
($28 \le f_{\rm \,NL} \le 48$), 
we then test an ensemble composed of CMB maps with contaminations of one of these 
types: half of them have primordial NG and the other half have secondary non-Gaussian 
contaminations. 
The results show that our method successfully classifies $\sim 95 \%$ of the tested 
maps as being Gaussian plus inhomogeneous noise CMB maps or CMB maps with 
primordial NG.

All these tests show that the method is very ef\/ficient when applied to data similar to which 
the NN was trained on. 
The best performance shown by our estimator occurs when the type of NG present in the data 
is known, so that the training of the NN with MCs contaminated with such NG is ef\/fective. 
This means that, when applied to data containing non-Gaussian components with which 
the NN have been not trained the estimator become imprecise. 
A similar situation was recently verified by~\cite{2013/modest}, who pointed out that simulated 
CMB maps, on large angular scales, with diverse levels of NG of local type cannot reproduce the 
anomalous outcomes found in the ILC-7yr full-sky WMAP map. 
Nevertheless, the results of our analyses of the foreground-cleaned {\it Planck} maps are in 
good agreement with last constraints, for large angular scales, from WMAP-9yr and 
{\it Planck} data analyses, i.e., $f_{\rm \,NL} = 38 \pm 18$~\citep{PLA-XXIV}. 
Summarizing, our conclusions point to the following approach: 
by considering a given type of contamination in the training data set, here the {\it local} NG, 
the NN recognizes satisfactorily $\gtrsim$ 98\% of the examined maps. 
This means that, adding contributions of other non-Gaussian types to simulated maps, in order 
to suitable instruct the NN's training data set, would let to an improved analysis of a CMB map. 
In this context, our results show that this estimator can be considered a useful tool to detect 
and constrain primordial, as well as secondary, non-Gaussian ef\/fects possibly present in CMB data.



\acknowledgments

We are grateful for the use of the Legacy Archive for Microwave Background Data 
Analysis (LAMBDA) and of the $\{ a^{\mbox{\footnotesize G}}_{\ell \, m} \}$ and $\{ a^{\mbox{\footnotesize NG}}_{\ell \, m} \}$ simulations \citep{Elsner}. 
We also acknowledge the use of CAMB (http://lambda.gsfc.nasa.gov/toolbox/tb\_camb\_form.cfm), developed by A. Lewis and A. Challinor (http://camb.info/)~\cite{CAMB}, 
and of the code for calculating the MFs, from \cite{Ducout} and \cite{2012/gay}.
Some of the results in this paper have been derived using the HEALPix 
package~\citep{Gorski}. 
AB acknowledges a CNPq fellowship, and CPN acknowledges the Capes and 
CNPq [237059/2012-6] fellowships. 
CAW acknowledges the CNPq grant 308202/2010-4.



\end{document}